\documentstyle[12pt,equation]{article}
\begin{document}

\newcommand{\tanb}{\mbox{$\tan \! \beta$}}
\newcommand{\msb}{\mbox{$m_{\tilde{b}_L}$}}
\newcommand{\mst}{\mbox{$m_{\tilde{t}_1}$}}
\newcommand{\mstt}{\mbox{$m_{\tilde{t}_2}$}}
\newcommand{\mstl}{\mbox{$m_{\tilde{t}_L}$}}
\newcommand{\mstr}{\mbox{$m_{\tilde{t}_R}$}}
\newcommand{\mstlr}{\mbox{$m_{\tilde{t}_{L,R}}$}}
\newcommand{\mstsq}{\mbox{$m^2_{\tilde{t}_1}$}}
\newcommand{\stst}{\mbox{$(\tilde{t}_1 \tilde{t}_1^*)$}}
\newcommand{\sigst}{\mbox{$\sigma_{\tilde{t}_1}$}}
\newcommand{\gamgam}{\mbox{$\gamma \gamma$}}
\newcommand{\st}{\mbox{$\tilde{t}_1$}}
\newcommand{\stt}{\mbox{$\tilde{t}_2$}}
\newcommand{\stl}{\mbox{$\tilde{t}_L$}}
\newcommand{\str}{\mbox{$\tilde{t}_R$}}
\newcommand{\ww}{\mbox{$W^+W^-$}}
\newcommand{\mat}{\mbox{${\cal M}^2_{\tilde{t}}$}}
\newcommand{\be}{\begin{equation}}
\newcommand{\ee}{\end{equation}}
\newcommand{\een}{\end{subequations}}
\newcommand{\ben}{\begin{subequations}}
\newcommand{\beq}{\begin{eqalignno}}
\newcommand{\eeq}{\end{eqalignno}}
\renewcommand{\thefootnote}{\fnsymbol{footnote} }
\noindent
\begin{flushright}
MAD/PH/795\\
KEK--TH--373\\
KEK Preprint 93--131\\
October 1993
\end{flushright}
\vspace{1.5cm}
\pagestyle{empty}
\begin{center}
{\Large \bf A new signal for scalar top bound state production} \\
\vspace{5mm}
Manuel Drees\footnote{Heisenberg Fellow}\\
{\em Physics Department, University of Wisconsin, Madison, WI 53706, USA}\\
Mihoko M. Nojiri\footnote{E--mail: NOJIRIN@JPNKEKVX}\\
{\em Theory Group, KEK, Oho 1--1, Tsukuba, Ibaraki 305, Japan}
\end{center}

\begin{abstract}
\noindent
We study the production and decay of a scalar \stst\ bound state \sigst\ at
hadron supercolliders, where \st\ is the lighter stop eigenstate. If \st\ has
no tree--level 2--body decays, the dominant decay modes of \sigst\ are $gg$
or, if $m_h < \mst \ll \mstt$, a pair of light scalar Higgs bosons $h$.
Nevertheless the branching ratio into two photons is often large enough to
yield a detectable signal.
\end{abstract}
\vspace*{1cm}
\clearpage
\noindent
\setcounter{footnote}{0}
\pagestyle{plain}
\setcounter{page}{1}
One of the best motivated extensions of the Standard Model (SM) of particle
physics is the introduction of supersymmetry (SUSY). On the one hand it
solves the ``hierarchy problem'', i.e. stabilizes the scale of electroweak
symmetry breaking against radiative corrections which would otherwise pull it
up to the unification scale $M_X$ or the Planck scale \cite{hierarchy}. On the
other hand recent precise measurements of the strength of coupling constants
exclude the possibility of Grand Unification if the only light (compared to
$M_X$) particles are those of the SM, while the minimal supersymmetric standard
model (MSSM) \cite{7} contains just the right degrees of freedom to allow for
unification at scale $M_X \simeq 10^{16}$ GeV \cite{gut}.

The MSSM predicts the existence of scalar superpartners to all known leptons
and quarks. It is by now quite well known that the lighter scalar top (stop)
eigenstate \st\ is expected to be lighter than the superpartners of the first
two generations of quarks, and might even be the lightest strongly interacting
sparticle \cite{1,2}. There are two reasons for this: Since the top quark is
heavy, $m_t \geq 113$ GeV \cite{cdf}, mixing between the superpartners \stl,
\str\ of left-- and right--handed top quarks cannot be neglected, in contrast
to the superpartners of light quarks. Furthermore, if we assume all squarks to
have the same mass at some very high (GUT, string or Planck) scale, radiative
corrections \cite{3} will reduce the masses of \stl\ and \str\ relative to
those of the other squarks.

Here we study possible signals for the production of a scalar \stst\ bound
state \sigst\ within the MSSM. Squark bound state production and decay were
first discussed in ref.\cite{rizzo}, for negligible mixing between the
superpartners of left-- and right--handed quarks. In ref.\cite{6} it has been
pointed out that this mixing can be very important for the case of \st\ bound
states; in particular, the branching ratio for $\sigst \rightarrow hh$ might
be large, where $h$ is the lighter scalar Higgs boson. Very recently it has
been claimed \cite{8} that $\sigst \rightarrow \ww$ can have a very large
branching ratio, which might give rise to interesting signals at hadron
supercolliders. We computed all potentially large branching ratios of \sigst.
We basically agree with the results of ref.\cite{6}, but were unable to
reproduce those of ref.\cite{8}. The cleanest signal for \sigst\ production at
hadron colliders arises from its 2--photon decay, giving rise to a peak in the
\gamgam\ invariant mass spectrum.

The starting point of our discussion is the stop mass matrix. Following the
convention of ref.\cite{9} we write it as (in the basis \stl, \str):
\be \label{e1}
\mat = \mbox{$ \left( \begin{array}{cc}
m^2_{\tilde{t}_L} + m_t^2 + 0.35 D_Z & - m_t (A_t + \mu \cot \! \beta) \\
- m_t (A_t + \mu \cot \! \beta ) & m^2_{\tilde{t}_R} + m_t^2 + 0.16 D_Z
\end{array} \right) $}. \ee
Here, $D_Z = M_Z^2 \cos\!2 \beta$ with $\tanb = \langle \bar{H^0} \rangle /
\langle H^0 \rangle$ as usual \cite{7}, $\mu$ is the supersymmetric Higgs(ino)
mass parameter, $A_t$ a trilinear soft supersymmetry breaking parameter, and
$m^2_{\tilde{t}_{L,R}}$ the nonsupersymmetric contributions to the squared
masses of the \stl, \str\ current states. The diagonalization of \mat\ is
straightforward. While the gauge interactions of \st\ only depend on the stop
mixing angle $\theta_t$, the couplings of stop squarks to Higgs bosons depend
on all parameters entering eq.(\ref{e1}); all these quantities therefore have
to be specified before \sigst\ branching ratios can be computed.

Both single stop decay and \stst\ annihilation contribute to \sigst\ decays.
In the first case either \st\ or $\tilde{t}_1^*$ decays, leaving the other
squark behind. We assume that the gluino is too heavy to be produced in these
decays. In general we then have to consider the following decay modes:
\be \label{e3}
\st \rightarrow b \tilde{W}_i, \ t \tilde{Z}_j, \ c \tilde{Z}_j,
\ i=1,2, \ j=1,\dots,4,
\ee
where $\tilde{W}_i$ ($\tilde{Z}_j$) denotes a generic chargino (neutralino)
state. The first two decays occur at tree level and with full gauge or top
Yukawa strength. It has been shown in ref.\cite{2} that $c \tilde{Z}_j$ is the
dominant \st\ decay mode if these first two modes are kinematically forbidden.
However, this last decay only occurs at 1--loop level and necessitates flavor
mixing; it is therefore suppressed relative to the tree--level decays by a
factor $|\epsilon|^2 \simeq 10^{-7}$ \cite{2}. We will see below that the $c
\tilde{Z}_j$ mode can therefore be neglected in the discussion of \sigst\
decays. The widths of (\ref{e3}) can be computed using the couplings of
refs.\cite{7}; the decay width of \sigst\ is twice that of \st.

In annihilation decays \st\ and $\tilde{t}_1^*$ annihilate into a flavor and
color singlet final state; this kind of decay is by far dominant for the
familiar $(c \bar{c})$ and $(b \bar{b})$ bound states (quarkonia). We
calculated the widths for the following modes:
\be \label{e4}
\sigst \rightarrow  gg, \ \ww, \ ZZ, \ Z \gamma, \ \gamgam, \
hh, \ b \bar{b}, \ t \bar{t}, \
\tilde{Z}_i \tilde{Z}_j, i,j=1,\dots,4.
\ee
Since \sigst\ is a scalar ($s-$wave) state, we only need \cite{10} the \st\
velocity $v \rightarrow 0$ limit of the $\st \tilde{t}_1^*$ annihilation
amplitudes leading to the final states of eq.(\ref{e4}). In this limit, the
$gg, \ \gamgam$ and $Z \gamma$ final states are produced via $t-$channel \st\
exchange as well as via 4--point ``butterfly'' interactions; $\ww$ is
accessible via $\tilde{b}_L$ exchange\footnote{We ignore mixing in the
$\tilde{b}$ sector.}, a 4--point interaction as well as scalar Higgs exchange
in the $s-$channel, while the $ZZ$ and $hh$ final states come from \st\ or
\stt\ exchange in the $t-$channel, a 4--point coupling and Higgs
exchange.\footnote{\stt\ exchange has not been included in ref.\cite{6}, where
the $hh$ final state has been studied; this contribution is small compared to
the \st\ exchange term for parameters leading to a sizable $hh$ branching
ratio.} The production of $Q \bar{Q}$ states involves $s-$channel scalar Higgs
exchange and $t-$channel chargino or neutralino exchange; note that the
corresponding matrix elements are proportional to the final state quark
masses, so that the $b \bar{b}$ width is very small unless $\tanb \gg 1$.
Finally, neutralino pairs can be produced via $s-$channel Higgs exchange or
$t-$channel top exchange.

In order to compute the decay widths for the processes (\ref{e4}) we have to
know the ``wave function at the origin'' $|\psi(0)|^2$, see ref.\cite{10}.
Recently the \mst\ dependence of this quantity has been parametrized in
ref.\cite{11}, for 4 different values of the QCD scale parameter $\Lambda$,
using a potential that reproduces the known quarkonium spectrum well; we use
their fit for $\Lambda = 0.2$ GeV.

In fig.1 we show examples for the branching ratios of the more important
processes of eq.(\ref{e4}) for relatively small $m_{\tilde{t}_{L,R}}=200$ GeV.
In addition to the parameters appearing in eq.(\ref{e1}) we have to specify
the SU(2) gaugino mass $M_2$ (we assume the usual unification condition $M_1 =
5/3 \tan^2 \theta_W M_2$) and the mass $m_P$ of the pseudoscalar Higgs boson.
This then determines all relevant masses and couplings. We have included
leading radiative corrections to the scalar Higgs sector involving top--stop
loops \cite{rad}.

In this figure we have assumed $M_2 = 100$ GeV leading to a mass of about 110
GeV for the lighter chargino. For $\mst > 115$ GeV the single stop decay into
$b+\tilde{W}_1$ (not shown) opens up and quickly dominates the total decay
width. Indeed, in this region the total width of \sigst\ is comparable to its
binding energy. Our calculations are no longer reliable in this case, since we
assume that formation and decay of \sigst\ occur at very different time scales
so that they can be treated independently; one has to use methods developed
previously \cite{12} for $(t \bar{t})$ bound states instead. However, we can
conclude from fig. 1 that if the tree--level single stop decays are allowed
the branching ratios for final states that might be detectable at hadron
colliders (see below) are very small, less than $10^{-4}$.

In fig. 2 we have therefore varied $M_2$ along with \mst, so that the decays
$\st \rightarrow b \tilde{W}_i, \ t \tilde{Z}_j$ remain closed for $\mst \leq
|\mu|$. We have also chosen larger values for $m_{\tilde{t}_{L,R}}$, with
$m_{\tilde{t}_L} > m_{\tilde{t}_R}$ as predicted by minimal supergravity
models \cite{9}. We see that now the $Br(\sigst \rightarrow hh)$ shoots up
very rapidly once this decay becomes kinematically allowed. The reason is that
the $h \st \st$ coupling \cite{7} contains a term which, in the limit $m_P^2
\gg M_Z^2$, is exactly proportional to the $LR$ element of \mat. Obviously
this element has to be large if \mst\ is to be much smaller than
$m_{\tilde{t}_{L,R}}$. As a result, the amplitude for $\st \tilde{t}_1^*
\rightarrow hh$ is proportional to $\left( m^2_{\tilde{t}_{L,R}}/M_W \mst
\right)^2$ if $m^2_{\tilde{t}_{L,R}} \gg \mstsq \gg m^2_h/2$. The
corresponding branching ratio therefore decreases with increasing \mst, in
spite of the increasing phase space.

The branching ratios for \ww\ and $ZZ$ also increase quickly just beyond
threshold. However, they do not reach the level of the $hh$ branching ratio;
their amplitudes are at best $\propto \left( m_{\tilde{t}_L}/M_W \right)^2$,
if $m^2_{\tilde{t}_{L,R}} \gg \mstsq \gg M^2_W$. This can be understood from
the equivalence theorem \cite{13}, which states that amplitudes involving
longitudinal gauge bosons are equal to corresponding ones involving
pseudoscalar Goldstone bosons $G$, if the energy of the process is $\gg m_W$.
There is no diagonal $G \st \st$ coupling; a $G \st \stt$ coupling with
strength similar to the $h \st \st$ coupling does exist, but it only affects
\sigst\ decays via diagrams involving a {\em heavy} stop propagator. The
amplitude for $\sigst \rightarrow Z_L Z_L$ is therefore suppressed by a factor
$\left( \mst / \mstt \right)^2$ compared to the $hh$ amplitude. Similar
arguments apply for the $W^+_L W^-_L$ amplitude. Transverse $W$ and $Z$ bosons
are at best produced with ordinary (weak) gauge strength, and their couplings
can even be suppressed by $\tilde{t}$ mixing. Unlike ref.\cite{8} we therefore
never find the width into $WW$ to exceed the one into gluons. However, the
authors of ref.\cite{8} neglected $\tilde{t}$ mixing, and assumed that
$m_{\tilde{t}_L}$ can be varied independently of the mass of the left--handed
sbottom $\tilde{b}_L$. (The same assumptions were made in ref.\cite{rizzo}.)
Since $\tilde{t}_L$ and $\tilde{b}_L$ reside in the same $SU(2)$ doublet, this
introduces a new source of explicit gauge symmetry breaking, which renders the
theory nonrenormalizable.

The structure of the curves of fig. 2 around \mst=250 GeV occurs because for
the given choice of parameters the $s-$channel heavy Higgs ($H$) exchange
diagrams become resonant here ($m_H \simeq 500$ GeV), greatly enhancing the
matrix elements for $t \bar{t}, \ b \bar{b}, \ hh$ and $\tilde{Z}_i
\tilde{Z}_j$ final states. The enhancement of the \ww\ and $ZZ$ channels
is much weaker, since the $HVV$ couplings ($V=W,Z$) are small for $m_H^2 \gg
M_Z^2$. If 2\mst\ is very close to $m_H$ our treatment again breaks down; in
this case the \stst\ bound states mix with $H$ \cite{14}. The curves of figs.
1,2 also exhibit numerous minima resulting from destructive interference
between different contributions to the matrix elements. Note that (at least in
the limit $v \rightarrow 0$) usually only a single partial wave is accessible
in \sigst\ decays; if in addition only a single combination of final state
helicities can be produced, destructive interference can lead to a vanishing
total amplitude even far above threshold.

Clearly one could in principle learn a lot about the MSSM parameters by
studying \sigst\ branching ratios. In practice, however, even discovery of
\sigst\ may not be trivial. We focus here on $pp$ supercolliders. The total
cross section for \sigst\ production at the SSC ($\sqrt{s}=40$ TeV) is shown
by the solid line in fig. 3. Hadronic final states ($gg, \ b \bar{b}, t
\bar{t}$) will be useless for the discovery of \sigst\ at hadron colliders,
due to the enormous backgrounds. In ref.\cite{8} the use of the \ww\ final
state was advocated. However, we have seen above that $SU(2)$ gauge invariance
implies a rather small rate for this final state; besides, it is not clear to
us how the \ww\ invariant mass will be measured, since both $W$ bosons will
have to decay leptonically in order to suppress QCD backgrounds. The $ZZ$
final state is very clean if both $Z$ bosons decay leptonically, but then the
rate will be very small even at the SSC ($<$ 5 events/year).

In ref.\cite{6} the use of $\sigst \rightarrow hh \rightarrow \tau^+ \tau^+
\tau^- \tau^-$ has been proposed. Since $Br(h \rightarrow \tau^+ \tau^-)
\simeq 8\%$ the 4 $\tau$ final state is also relatively rare ($Br < 6.4
\cdot 10^{-3}$). However, it is not clear how this final
state is to be identified experimentally in a hadronic environment. A pair
of isolated like--sign leptons would be a rather clean tag; assuming a $30\%$
detection efficiency per lepton one could get more than 50 events/year for
$\mst \leq 100$ GeV if the $hh$ final state dominates. The problem is that
the presence of at least 4 neutrinos in the final state makes it impossible to
reconstruct the \sigst\ mass. Given that there are sizable backgrounds
(e.g., $\sigma(pp \rightarrow ZZX \rightarrow  \tau^+ \tau^+
\tau^- \tau^- X) \simeq 3 \cdot 10^{-2}$ pb, leading to $\sim 300$
events/year) the identification of the 4 $\tau$ signal might be problematic.

Probably the most promising signal results from the decay $\sigst \rightarrow
\gamgam$. Searches for intermediate mass Higgs bosons will presumably require
good electromagnetic energy resolution for SSC detectors, so the
reconstruction of $m(\sigst)$ is in principle straightforward. The signal
would then be a bump in the \gamgam\ spectrum on top of the smooth
background. In order to suppress QCD backgrounds (misidentified $\pi^0$s)
as well as backgrounds from photon bremsstrahlung off quarks one usually
requires the signal photons to be isolated, i.e. to have little hadronic
activity in a cone around the photon. In addition, the physics background
from $q \bar{q}$ annihilation and $gg$ fusion is strongly peaked in forward
and backward directions, due to propagator effects; it is therefore
advantageous to demand the centre--of--mass scattering angle $\theta^*$ of
the two photons to be large \cite{backgr}. We impose the
following cuts:
\ben \label{e5} \beq
|\cos \! \theta^*| \ & \leq 0.5 ; \label{e5a} \\
|y_{\gamma}| &\leq 1.74. \label{e5b}
\eeq \een
The cut on the rapidity $y_{\gamma}$ guarantees that the photons are well
isolated from the beam pipes (lab scattering angle $\theta \geq 20^{\circ}$).
It also helps to further suppress the $q \bar{q}$ background, which comes from
an asymmetric initial state (valence quark on sea antiquark) and hence often
undergoes a strong boost.

The dashed curves in fig. 3 show the two--photon signal for the two sets of
parameters shown in figs. 1 and 2, respectively, after the cuts of
eqs.(\ref{e5}) have been applied. Clearly there is a sizable number of signal
events for almost all combinations of parameters, as long as tree--level
2--body decays of \st\ are suppressed. In order to decide whether this signal
can also be seen on top of the irreducible (physics) background we have
defined the minimal cross section leading to a significant signal after one
year of SSC running (integrated luminosity = $10^4$ pb$^{-1}$). Following
ref.\cite{15} we have defined the signal to be significant if the 99\% c.l.
lower limit of signal + background (after cuts) is at least as large as the
99\% upper limit of the background alone. The background has been integrated
over an invariant mass bin of width $0.02 \cdot M_{\gamgam}$, which requires
good energy resolution. The resulting minimal detectable signal is shown
by the dotted line in fig. 3. This line scales up with the square--root of the
bin width over which the background has to be integrated, and scales down with
the square--root of the integrated luminosity. We conclude from this figure
that \sigst\ should be quite easily detectable at the SSC if tree--level \st\
decays are suppressed, the $hh$ channel is not greatly enhanced by very large
off--diagonal elements of the stop mass matrix, and if one is not too close to
the $H$ pole. The long--dashed curve shows that even in the presence of a
sizable branching ratio into the $hh$ final state the signal might still be
visible after several years of operation, provided the energy resolution of
the calorimeters is not much worse than we assumed. We mention in passing that
the superpartners of light quarks do usually have unsuppressed 2--body decays;
we do therefore not expect their bound states to be detectable at the SSC.

We computed both signal and background in leading order in QCD, since no
higher order calculation of the signal rate is available yet. Such
calculations do exist for the background \cite{nlo}. The result is very
similar to the leading order prediction if, in addition to the cuts
(\ref{e5}), one vetoes against the presence of high--$p_T$ jets.

In summary, we have computed all potentially large branching ratios of a
scalar stoponium bound state \sigst, fully taking into account effects due to
$\tilde{t}$ mixing. The dominant decay modes are $gg, \ hh$
or, if $m(\sigst) \simeq m_H, \ t \bar{t}$. The process $pp \rightarrow
\sigst X \rightarrow \gamgam X$ should be observable at the SSC,
{\em provided} that \st\ has no unsuppressed tree--level 2--body decays.

\subsection*{Acknowledgements}
We thank Howard Arthur Baer for providing us with a program for the
computation of the $\gamgam$ physics background, X. Tata for useful
discussions and suggestions, and H. Inazawa and T. Morii for communications on
ref.\cite{8}. The work of M.D. was supported in part by the U.S. Department of
Energy under contract No. DE-AC02-76ER00881, by the Wisconsin Research
Committee with funds granted by the Wisconsin Alumni Research Foundation, by
the Texas National Research Laboratory Commission under grant RGFY93--221, as
well as by a grant from the Deutsche Forschungsgemeinschaft under the
Heisenberg program.

\renewcommand{\theequation}{A.\arabic{equation}}
\setcounter{equation}{0}

\clearpage
\section*{Figure Captions}

\renewcommand{\labelenumi}{Fig.\arabic{enumi}}
\begin{enumerate}

\item 
Branching ratios for annihilation decays of \sigst\ listed in eq.(\ref{e4}).
The range of \mst\ values shown results from varying $A_t$ between --310 and
--70 GeV. The values of the other parameters are: $\mstl=\mstr=200$ GeV,
$m_t = -\mu = 150$ GeV, $M_2 = 100$ GeV, $m_P = 500$ GeV and \tanb\ = 2.
The branching ratios for the $b \bar{b}$ and $t \bar{t}$ final states (not
shown) are always below 10$^{-3}$.

\vspace*{5mm}
\item   
Branching ratios for annihilation decays of \sigst\ listed in eq.(\ref{e4}).
The range of \mst\ values shown results from varying $A_t$ between 440 and
1080 GeV. We have increased the $SU(2)$ gaugino mass $M_2$ along with \mst\,
so that the tree--level single stop decays of eq.(\ref{e3}) remain
kinematically forbidden ($M_2 = 1.5 \mst$). The values of the other parameters
are: $m_t = 150$ GeV, \mstl\ = 400 GeV, \mstr\ = $-\mu=$ 300 GeV, $m_P = 500$
GeV, and \tanb\ = 2. The branching ratios for the $b \bar{b}$ and $t \bar{t}$
final states (not shown) are again small, except for the vicinity of the $H$
pole ($\mst \simeq 250$ GeV), where the $t \bar{t}$ final state dominates.

\vspace*{5mm}
\item  
Cross section for \sigst\ production at the SSC. The solid line shows the
total cross section multiplied with 0.01, and the dashed curves the $\gamgam$
signal cross section after cuts, for the two scenarios of figs. 1 and 2. The
dotted curve shows the minimal cross section giving a significant signal after
one year of nominal SSC operations, as defined in the text.

\end{enumerate}

\begin{thebibliography}{99}
\bibitem{hierarchy}
E. Witten, Nucl. Phys. {\bf B188}, 513 (1981).

\bibitem{7}
H.E. Haber and G.L. Kane, Phys. Rep. {\bf 117}, 75 (1985); J.F. Gunion and
H.E. Haber, Nucl. Phys. {\bf B272}, 1 (1986).

\bibitem{gut}
U. Amaldi, W. de Boer and H. F\"urstenau, Phys. Lett, {\bf B260}, 447 (1991);
P. Langacker and M. Luo, Phys. Rev. {\bf D44}, 817 (1991); J. Ellis, S.
Kelley and D.V. Nanopoulos, Phys. Lett. {\bf B260}, 131 (1991).

\bibitem{1}
J. Ellis and S. Rudaz, Phys. Lett. {\bf B128}, 248 (1983).

\bibitem{2}
K.--I. Hikasa and M. Kobayashi, Phys. Rev. {\bf D36}, 724 (1987).

\bibitem{cdf}
H. Frisch, talk presented at the {\it 23rd International Symposium on
Multiparticle Dynamics}, Aspen, Colo., September 1993.

\bibitem{3}
K. Inoue, A. Kakuto, H. Komatsu and S. Takeshita, Prog. Theor. Phys. {\bf 68}
927 (1982); {\bf 71}, 413 (1984).

\bibitem{rizzo}
M.J. Herrero, A. M\'endez and T.G. Rizzo, Phys. Lett. {\bf B200}, 205
(1988).

\bibitem{6}
V. Barger and W.--Y. Keung, Phys. Lett. {\bf B211}, 355 (1988).

\bibitem{8}
H. Inazawa and T. Morii, Phys. Rev. Lett. {\bf 70}, 2992 (1993).

\bibitem{9}
M. Drees and M.M. Nojiri, Nucl. Phys. {\bf B369}, 54 (1992).

\bibitem{10}
V. Barger, E.W.N. Glover, K.--I. Hikasa, W.--Y Keung, M.G. Olsson, C.J.
Suchyta and X.R. Tata, Phys. Rev. {\bf D35}, 3366 (1987).

\bibitem{11}
K. Hagiwara, K. Kato, A.D. Martin and C.--K. Ng, Nucl. Phys. {\bf B344}, 1
(1990).

\bibitem{rad}
Y. Okada, M. Yamaguchi and T. Yanagida, Prog. Theor. Phys. {\bf 85}, 1 (1991),
and Phys. Lett. {\bf B262}, 54 (1991); H.E. Haber and R. Hempfling,
Phys. Rev. Lett. {\bf 66}, 1815 (1991); J. Ellis, G. Ridolfi and F. Zwirner,
Phys. Lett. {\bf B257}, 83 (1991), and {\bf B262}, 477 (1991).

\bibitem{12}
V.S. Fadin and  V.A. Khoze, JETP Lett. {\bf 46}, 525 (1987);
M.J. Strassler and M.E. Peskin, Phys. Rev. {\bf D43},  1500 (1991);
Y. Sumino, K. Fujii, K. Hagiwara, H. Murayama and C.--K. Ng, Phys. Rev.
{\bf D47}, 56 (1992).

\bibitem{13}
B.W. Lee, C. Quigg and H. Thacker, Phys. Rev. {\bf D16}, 1519 (1977); M.S.
Chanowitz and M.K. Gaillard, Nucl. Phys. {\bf B261}, 379 (1985); H. Veltman,
Phys. Rev. {\bf D41}, 2294 (1990).

\bibitem{14}
J. Ellis, M.K. Gaillard, D.V. Nanopoulos and C.T. Sachrajda, Phys. Lett.
{\bf B83}, 339 (1979); M. Drees and K.--I. Hikasa, Phys. Rev. {\bf D41},
1547 (1990).

\bibitem{backgr}
D.A. Dicus and S.S.D. Willenbrock, Phys. Rev. {\bf D37}, 1801 (1988);
H.A. Baer and J.F. Owens, Phys. Lett. {\bf B205}, 377 (1988).

\bibitem{15}
H.A. Baer, M. Bisset, C. Kao and X.R. Tata, Phys. Rev. {\bf D46}, 1067 (1992).

\bibitem{nlo}
B. Bailey and J.F. Owens, Phys. Rev. {\bf D47}, 2735 (1993).

\end{thebibliography}
\end{document}